\documentclass[12pt, natbib, preprint, a4paper]{aastex}

\newcommand{\be}{\begin{equation}}
\newcommand{\bb}{{{\mbox{\boldmath$B$}}}} 
\newcommand{\bs}{\begin{subequations}}
\newcommand{\cm}{{\:\rm cm}}

\newcommand{\ee}{\end{equation}}
\newcommand{\es}{\end{subequations}}
\newcommand{\erg}{{\:\rm erg}}
\newcommand{\ergs}{{{\:\rm erg}\; \rm s^{-1}}}
\newcommand{\g}{{\:\rm g}}
\newcommand{\G}{{\:\rm G}}
\newcommand{\hr}{{\:\rm hr}}
\newcommand{\jj}{{{\mbox{\boldmath$J$}}}}
\newcommand{\jorb}{{J_{\rm orb}}}
\newcommand{\kk}{{\widehat{\mbox{\boldmath{$k$}}}}} 

\newcommand{\muns}{{\mbox{\boldmath{$\mu$}}_1}}
\newcommand{\murd}{{\mbox{\boldmath{$\mu$}}_2}}
\newcommand{\ms}{{\:\rm ms}} 
\newcommand{\msun}{{\:M_\odot}} 
\newcommand{\n}{{\widehat{\mbox{\boldmath{$n$}}}}} 
\newcommand{\nn}{{{\mbox{\boldmath$N$}}}}
\newcommand{\porb}{{P_{\rm orb}}}
\newcommand{\s}{{\:\rm s}}
\newcommand{\yr}{{\:\rm yr}}

\usepackage{amssymb}                
\usepackage{rotating}               
\usepackage{amsfonts}

\shorttitle{A Magnetar in  a Young Low Mass Binary?}
\shortauthors{Pizzolato et al.}
 
\begin{document}


\title{1E161348-5055 in the Supernova Remnant RCW~103: 
A Magnetar in a  Young Low Mass Binary System?}


\author{
Fabio Pizzolato\altaffilmark{1}, 
Monica Colpi\altaffilmark{1},
Andrea De Luca\altaffilmark{2},
Sandro Mereghetti \altaffilmark{2},
Andrea Tiengo\altaffilmark{2}
} 

\altaffiltext{1}{Department Of Physics, University of Milano--Bicocca,
Piazza della Scienza 3 20126 Milano -- Italy}
\altaffiltext{2}{Istituto Nazionale di Astrofisica--Istituto di Astrofisica Spaziale  e Fisica Cosmica, Via Bassini 15 20133 Milano -- Italy} 

\email{fabio.pizzolato@mib.infn.it}


\begin{abstract}

We suggest that
the unique X-ray source 1E161348-5055 at the centre of 
the supernova remnant RCW~103 consists of a  neutron star 
in close orbit with a low mass main sequence star. 
The time signature of $6.67\hr$ is interpreted as the 
neutron star's spin period.  This requires the 
neutron star to be endowed  with  a high surface magnetic field 
of~$\sim 10^{15}\G$. 
Magnetic or/and material (propeller) torques 
are able to spin rapidly the young neutron 
star down to an asymptotic, equilibrium  
spin period in close synchronism with the orbital period, 
similarly to what happens in the
Polar Cataclysmic Variables.
1E161348-5055 could be the first case of
a magnetar born in a young low mass binary system.

\end{abstract}




\maketitle


\section{Introduction}
\label{s-intro}

The soft X-ray source 1E 161348-5055 (hereafter 1E)
was discovered by \citet{Tuo80} with the {\sl Einstein} 
observatory, close to the
geometrical centre of the young supernova remnant (SNR)  RCW~103. 
Owing to its thermal-like spectrum and
to the lack of a counterpart at radio or optical
wavelengths, 1E was classified as the first
example of a radio-quiet, isolated neutron star.
The host SNR is very young \citep[$\sim2000\yr$,][]{Car97}
and it is located at a distance of $\sim3.3~\rm kpc$
\citep{Cas75,Rey04}. The association of 1E
and RCW~103 is very robust. The point source
lies within $15''$ of the SNR centre;
moreover,  1E and RCW~103 have consistent 
distance measurements \citep{Rey04}.

The original interpretation of 1E as an isolated neutron
star was later questioned by the observation
of a large variability on few years'  time scale
\citep{Got99}. This was confirmed 
in recent years, when a factor $\sim100$
brightening (from $\sim9\times 10^{-13} \erg \cm^{-2}\s^{-1}$
to $\sim7\times 10^{-10}\erg \cm^{-2}\s^{-1}$)
was discovered, comparing the first
and the second {\sl Chandra} observations of the source, performed in 
September 1999 and in March 2000, respectively \citep{Gar00a}.
Even more puzzling, a possible long periodicity at $\sim6$ hours
was hinted by the first {\sl Chandra} observation \citep{Gar00b}, but
it was not recognised in the second {\sl Chandra} data set.
Subsequent observations with {\sl Chandra} \citep{San02} and 
{\sl XMM-Newton} \citep{Bec02} failed to ultimately settle the periodicity issue.

A breakthrough came with the deep ($90~\rm ks$) observation of 1E 
performed by {\sl XMM-Newton} in 2005. \citet{DeL06} caught the source
in a low state 
($1.7 \times10^{-12} \erg\cm^{-2}\s^{-1}$)
and reported conclusive evidence for a strong
($\sim50\%$), nearly sinusoidal modulation of 1E at $P=6.67\pm0.03\hr$.
The source spectrum, thermal-like and well described by the sum of two
black bodies, varies along the $6.67\hr$
cycle and is harder at the pulse maximum \citep{DeL06}.
The $6.67\hr$ periodicity was also recognised in the former 
{\sl XMM-Newton} data set (collected in 2001, when the source was a factor 
$\sim5$ brighter),
although with a much smaller pulsed fraction ($\sim12\%$) and
with a remarkably different, complex light curve, featuring two 
narrow minima (or ``dips'') per period, separated by $0.5$ in phase.
No faster periodicities were found, with a $3~\sigma$
upper limit of $10\%$  on the pulsed fraction down to $P=12~\rm ms$.

\citet{DeL06} reported also a complete picture of the 
peculiar long-term variability of 1E. The source has been continuously
fading along more than $5$ years,
since the 1999-2000 re-brightening.
In August 2005 1E was seen with a flux similar to
the pre-outburst level.
{\sl XMM-Newton}  data clearly
show that the spectrum of the source is also evolving. 
While a double black body model yields in any case the best 
description of the spectrum, the emission is
harder and more absorbed when the source is brighter. 

On the optical side, deep observations of the field with the 
{\sl ESO/VLT} and with {\sl HST} showed nothing but two/three very faint IR sources ($H\sim22$)  consistent with the accurate {\sl Chandra} position
\citep[][and reference therein]{DeL06}. No firm conclusions
could be drawn about the possible association of any of such
sources with 1E.

\subsection{Isolated vs Double Star Scenario}

The combination of long-term variability, 6.67 hours periodicity,
young age and under-luminous optical/IR counterpart makes 1E a 
unique source among compact objects in SNRs. 

As discussed by \citet{DeL06}, 1E could either be a very young
low mass X-ray binary (LMXB) system, featuring a $2000\yr$ old 
compact object and
a low mass companion star (M4, or later) in an eccentric orbit,
or a very peculiar isolated neutron star. In the latter case,
1E could only be a  magnetar, dramatically slowed 
down, possibly by a propeller interaction with a debris disc.
Both scenarios require highly non-standard assumptions.

The binary scenario is appealing since it provides
an easy explanation of the long $6.67 \hr$ periodicity, as
due to the orbital period of the system. However, the observed properties
of 1E are dramatically different from those of any known LMXB. The system
is orders of magnitude dimmer than persistent LMXBs. 
Moreover, a very peculiar
``double accretion'' mechanism is required to explain the unusual,
large flux and spectral variations along the $6.67 \hr$ cycle
as an orbital modulation, as well as the dramatic long term variability.  
Indeed, wind accretion 
along an eccentric orbit could naturally explain the quasi-sinusoidal
modulation observed in the low state.
However, the light curve of the source in its  ``active state''
is very complex. In order to explain its features within
such a scenario - as due to varying occultation of the
central source - the existence of very peculiar structures 
at the rim of a (possibly transient) accretion disc 
is required.

The isolated magnetar scenario could easily explain 
the phenomenology of the periodic modulation, with the
star rotation bringing into view or hiding different emission regions.
Indeed, the light curve profile is very similar to the one observed
in most magnetars.
The long-term variability (both in luminosity and in pulse profile)
is also reminiscent of the one seen in
a few ``transient'' anomalous X-ray pulsars (AXPs).
However, all known magnetars spin thousands of time faster than 1E, with
periods well clustered in the $5-12 \s$ range. The spin history of 
1E should have been dramatically different.
\citet{DeL06}  and \citet{Li07} 
showed that propeller interaction with a supernova 
debris disc could provide a very efficient slowing-down mechanism.
Such a mechanism could brake the  neutron star to $6.67\hr$ 
in $2000\yr$, provided that
the star was born with a $\sim10^{15}\G$ magnetic field and an initial
period not faster than $300~\rm ms$, in order
to avoid an initial "ejector" phase which would have pushed away
any surrounding material \citep[see e.g.][]{Lip92}. 
Indeed, evidence for a debris
disc surrounding the AXP $4U~0142+61$
has been recently obtained \citep{Wan06}. However, this source
spins at $P=8.4\s$, a very typical value for an AXP.
Thus, peculiar birth properties, coupled to some fine tuning,
should be required to explain the unique spin evolution of 1E in the
isolated magnetar scenario.

In this paper we will explore an alternative scenario, also suggested by
\citet{Pop06} as a possibility,  for the 
nature of 1E. We assume that 1E is a binary system,
hosting a recently born neutron star and a low mass companion. 
Within such a frame, we will suppose 
the $6.67\hr$ periodicity to be the spin period 
of the neutron star. The required spin-down  mechanism 
is provided by material and/or magnetic interaction between the 
neutron star and its stellar companion. 
A magnetar-like field of~$\sim10^{15}\G$ is required  for these
interactions to brake the neutron star's rotation that slow.

The binary system  envisaged in our model has some interesting 
analogies with some magnetic Cataclysmic Variables. Indeed,
the present investigation is partially motivated by the similarity,
already pointed out by \citet{DeL06},
of several temporal properties of 1E (the period value, the pulse shape
and its long-term variability) with those  of Polars
(also known as {\sl AM Herculis}), 
a class of strongly magnetised
Cataclysmic Variables (CVs; see e.g. \citealp{Cro90} for a review).

In such systems the magnetic interaction between the components is
so strong to lock the white dwarf in synchronous rotation with the 
orbit, with periods  up to few hours (the longest period of a  Polar,  
{\sl V1309~Ori}, is of $8\hr$, \citealp{Gar94}). 
In addition, the magnetic pressure prevents 
the formation of a disc around the white dwarf, and  
the accretion flow is directly  channelled
from the inner Lagrangian point $L_1$ to one or
both the  white dwarf's poles.
Our model  owes a good deal of inspiration
to the current theoretical  models of strongly magnetic CVs.
The reader is warned, however, that the ``Polar''  model we envisage for 1E
may differ in  some important details from  its CV counterpart.

We shall address the fundamentals of the model, deferring to a
further paper the detailed explanation of the light curve and the 
spectral properties of the source.

In Section~\ref{s-setup} we elaborate more quantitatively on the ``Polar'' 
model. 
In  Section~\ref{s-dynamics} we discuss the dynamical equations which
rule the system's evolution  after the 
supernova explosion which left the SNR RCW~103. 
In  Section~\ref{s-polar} we solve numerically these equations,
and  briefly discuss the properties of
their asymptotic equilibrium solutions (this topic is covered more fully in
Appendix~\ref{s-equilibrium}). 
In Section~\ref{s-discussion} we cover a number of important issues such as
the secondary's evaporation  and the future evolution of
1E. Our  conclusions are  summarised in Section~\ref{s-summary}.

\section{Setting up  the Model}
\label{s-setup}

We assume that 1E is a binary system made by a neutron 
star (NS) of mass $M_1=1.4\msun$ and a low main sequence star (a red dwarf, RD)
of mass  $M_2=0.4\msun$ (consistent with the current 
optical/infrared constraints). 
This system survived the supernova explosion which
led to the birth of RCW~103. The orbit is expected to be born quite
eccentric, with $e$ most likely in the range $0.2-0.5$
\citep{DeL06}; this eccentricity
deserves some comments.
In a narrow system like that we envisage for 1E, the tidal torques are
expected to be quite effective in 
synchronising the companion star's rotation with the orbital period and in circularising the orbit.
Unfortunately, there is no general consensus about the
efficiency of these two processes (see e.g. \citealp{Zah77}; \citealp{Tas88};
\citet{Rie97}; \citealp{Tas97}; see also \citealp{Mei05} and references therein).
The most conservative estimates of the synchronisation and circularisation
time scales are those derived by \citet{Zah77} (his Equations (4.12) and
(4.13))
\begin{eqnarray}
\label{e-tides}
t_{\rm sync} = \frac{1}{6\; k_2\; q^2}\; \left(\frac{M_2\; R_2^2}{L_2}\right)^{1/3} \; \frac{I_2}{M_2\; R_2^2}\;
\left(\frac{a}{R_2}\right)^6
\\
t_{\rm circ} = \frac{1}{63\; q(1+q)\; k_2}\;\left(\frac{M_2\; R_2^2}{L_2}\right)^{1/3} \; \left(\frac{a}{R_2}\right)^8,
\end{eqnarray}
where $L_2\simeq 2\times 10^{32}\ergs$ is the main sequence star's luminosity, 
$q=M_1/M_2$ is the mass ratio,
$I_2$ and $k_2$ are the main sequence star's momentum of inertia and the
apsidal motion constant (see \citealp{Zah77} for more details).
The quantities $k_2$ and $I_2/M_2\,R_2^2$ are both  of the same 
order (see e.g. \citealp{Mot52}), 
and therefore should approximately cancel each other, so  
a stellar system with $P_{\rm orb}=6.67\hr$, $M_1=1.4\msun$ and $M_2=0.4\msun$
would take $t_{\rm sync}\simeq 10^2\yr$ to synchronise and
$t_{\rm circ}\simeq 2\times10^3\yr$ to circularise.  The companion star is thus
expected to be synchronous, but the orbit may still be eccentric.

Although the eccentricity may contribute to modulate
the X--ray pulsations of 1E, its introduction
at the present level of discussion
would complicate the model without any real benefit. 
As  it will be discussed in Section~\ref{s-discussion}, our results are
not expected to be too  sensitive  to the use of a circular
orbit approximation.  
For this reason, we suppose that the orbit is
circular, with the two stars separated by the fixed distance  $a$.
If the NS's period $P_1=6.67\hr$ is equal (or close to)  the orbital period 
$P_{\rm orb}$, $a$ is given by Kepler's third law
\be
\label{e-axis}
a = 1.5 \times 10^{11} \; 
\left(\frac{\porb}{6.67\hr}\right)^{2/3}
\left(\frac{M}{1.8\msun}\right)^{1/3}
\cm,
\ee
where $M=M_1+M_2$ is the binary's total mass.

We assume that the interaction between the stars and the accretion flow are
dominated by the star's magnetic field, as happens in canonical
Polar CVs. 
This requires the condition  $R_A\gtrsim a$, where 
\be
\label{e-alfven}
R_A = 1.8\times 10^{11} \; 
\left(\frac{\mu_1}{10^{33}\G\cm^3}\right)^{4/7} \;
 \left(\frac{\dot{M}_1}{10^{13}\g\s^{-1}}\right)^{-2/7} \;
\left(\frac{M_1}{1.4\msun}\right)^{-1/7}\;\cm,  
\ee 
is the NS Alfv\'en radius. The mass accretion rate $\dot{M}_1$  has been  
normalised to the low state X-ray luminosity $L_X$ of 1E according to the 
well-known formula $L_X = G \: M_1\; \dot{M}_1/R_1$,
being $R_1 = 10^6\cm$ the NS radius.
It is easy to check that the requirement
$R_A\gtrsim a\simeq 10^{11}\cm$ implies a NS  magnetic moment  
$\mu_1\simeq 10^{33}-10^{34}\G\cm^3$, or  
a magnetar-like surface magnetic field 
$B_1\simeq \mu_1/R_1^3\simeq 10^{15}-10^{16}\G$.

The Roche lobe  of the NS is entirely filled by the 
star's own magnetosphere, and under this circumstance the formation of
an accretion disc is impossible: any mass flow 
from the RD  through $L_1$ is 
immediately channelled along the NS's field lines  and accreted.
Independently of  the accretion pattern, the balance between the 
material accretion spin up torque and the magnetic spin down
propeller torque fixes the equilibrium condition  $R_{\rm co}\sim R_A$
between the Alfv\'en radius and the co-rotation radius
\be
\label{e-rco}
R_{\rm co} =\left(G\, M_1/\omega_1^2\right)^{1/3},
\ee
where $\omega_1=2\,\pi/P_1$ is the NS rotation angular frequency. 
The rotation equilibrium condition $R_{\rm co}\sim R_A$  
sets the NS final  spin period
\be
\label{e-peq}
P_1 \to
P_{\rm eq} \simeq 6\; 
\left(\frac{\mu_1}{10^{33}\G\cm^3}\right)^{6/7} \;
 \left(\frac{|\dot{M}_2|}{3\times10^{13}\g\s^{-1}}\right)^{-3/7} \;
\left(\frac{M_1}{1.4\msun}\right)^{-5/7}\;\hr. 
\ee

In a Polar system (and also in 1E) the RD is endowed with a large magnetic 
moment. This may be 
either {\em induced} by the compact star or {\em intrinsic}. 
This last alternative is quite possible, as
it is known that M dwarf stars may harbour sizeable 
magnetic fields, even of few~$\rm kG$ 
(see e.g. \citealp{Don06}), and magnetic moments above $10^{34}\G\cm^3$.
In this paper we assume for the RD the  magnetic moment $\mu_2=10^{34}\G\cm^3$.

The  X-ray emission from Polars is powered by mass accretion from the secondary.
In our case, this may still be true, but some important contributions
may come from other sources as well.
Currently known magnetar candidates (Anomalous X--ray Pulsars, AXPs 
and Soft Gamma-ray Repeaters, SGRs: see \citealp{Woo06} 
for a review) are observed to emit X--ray radiation with luminosities
ranging from~$\sim 10^{33}$ to~$\sim 10^{36}\ergs$. This
emission is believed to be powered by the decay of their ultra-strong
magnetic field; this mechanism might be also responsible for the
$\sim 10^{33}-10^{35}\ergs$ X-ray luminosity of 1E. In
addition, the spectrum and the long term variability of 1E are remarkably
similar to the characteristics of some AXPs and SGRs (for example, the AXP
XTE~J1810--197, \citealp{Got05}).
Therefore, the X-ray emission of 1E might be (completely or
partially) explained by typical magnetar emission rather than by
accretion, while the main role of the companion star invoked in this paper
is to explain the discrepancy between the $6.67\hr$ pulsation period of 1E
and the much shorter periods $5-12\s$ observed in AXPs and SGRs.
If this is the case, the X-ray luminosity $L_X$ only provides an upper
limit to the mass accretion rate $\dot{M}_1$  on the NS: in the low
state of 1E, for instance, $\dot{M}_1 \lesssim 10^{13} \g\s^{-1}$.
Of course, the companion star may transfer mass to the compact object  
only provided that its size is of the same size as its Roche lobe.
According to the similarity relation valid for low ZAMS stars 
\citep[e.g.][]{Kip94}  the radius of the companion star is
\be
\label{e-lms}
R_2 = 3.3\times 10^{10}\left(\frac{M_2}{0.4\msun}\right)^{0.8}\cm,
\ee
and is comparable to the star's Roche lobe radius~$R_L\sim 4\times 10^{10}\cm$,
as computed from Equation~(2) of \citet{Egg83}.
Since $R_2$ is close to $R_L$, in view of the uncertainties, we need to explore
both the possibilities that the system is detached ($R_2<R_L$)
or that there is a Roche lobe overflow from the RD to the NS ($R_2\gtrsim R_L$).
The geometry of the system (i.e. detached or not), 
plays  a fundamental role  in the NS spin evolution. 
Indeed, Polars may reach exact synchronism 
provided that the magnetic torque is stronger than the accretion spin-up 
torque~$\propto \dot{M}_1$ \citep[e.g.][]{Cam86a}. 
If most of the  X-ray luminosity of 1E is 
powered by magnetar emission and not by accretion, $\dot{M}_1$  is small
(the binary may be detached), so the magnetic torques are dominant and may synchronise the system.
Note that in order to attain synchronism 
the magnetic torque  must not (only) be 
stronger the accretion torque {\em now}, 
but it must have been so also over a significant fraction 
if the system's life.
Incidentally, we note that if deeper observations will show the mass of the companion to be  significantly smaller than its current upper limit 
($0.4\msun$), then the companion is
most likely to lie well within its own Roche lobe. In this case the system
is detached, and the X-ray luminosity is clearly due to the  magnetar  emisson.
On the other hand,  if the  accretion torque is strong enough, it  may  keep  the NS away from  exact synchronism. In this case 1E
should be more similar to an  Intermediate Polar (IP).
The spin evolution of these systems is dominated by the material
torques,  and their action leads to rotation equilibrium 
periods distributed over a fairly wide range of values, all shorter
than the orbital period \citep{Nor04}.

In the next Section we set up all the formal machinery to deal with both 
these cases, aiming to follow the spin evolution of 1E.

\section{Dynamics of the System}
\label{s-dynamics}

In this Section we focus on the  evolution of the NS spin 
and the orbital period after the supernova explosion. 
In order to keep  our analysis as simple as possible, we
assume that all the torques are directed perpendicular to the 
orbital plane.  Any misalignment would lead to  a rotation of the
orbital angular momentum: this just complicates the equations, without any
relevant benefit for the physical insight into  the problem.

We evaluate the torques  acting on the NS, as well as
those acting on the orbit and the RD, as they may
redistribute angular momentum among  the different components.

\subsection{Torques acting on the  NS}
\label{s-ns}

The torques acting on the NS are the following.

\begin{enumerate}

\item 
The spin-down torque due to the  magneto-dipole losses \citep{Lip92} 
\be
\label{e-md}
\nn_{\rm md} =  - \twothirds \; 
 \frac{\mu_1^2 \;  \omega_1^3}{c^3} \;  \sin^2\chi \;
\kk_1
\ee
being $\kk_1$ the direction of the NS spin,
$\chi$ is the angle between the magnetic and the spin axes,
and  $\mu_1 = B_1\; R_1^3/2$ is  the NS magnetic moment's modulus.
In the following we assume $\kk_1$ 
coincident with  the unit vector $\kk$  
perpendicular to the orbital plane,  and
$\chi=\pi/2$, i.e. the NS's magnetic moment
$\muns$ lies in the orbital plane.

We may assign a characteristic time scale to the torque~(\ref{e-md})
\be
\label{e-time-md}
t_{\rm md} = \frac{\omega_1 \; I_1}{|\nn_{\rm md}|}
\simeq
3 \times 10^{-3}  \;
\left(\frac{I_1}{10^{45}\g \cm^2}\right) \;
\left(\frac{\mu_1}{10^{33}\G\cm^3}\right)^{-2} \;
\left(\frac{P_1}{10\ms}\right)^2\;\yr,
\ee
where $P_1$  and $I_1$ are  NS spin period and  moment of inertia.
On account  of the strong dependence $t_{\rm md}\propto P_1^2$,
 this torque is relevant in the early
NS life, when $P_1$ is small. It is fundamental in
braking the initially fast NS  rotation, but it  soon becomes
negligible and does not affect the final equilibrium rotation period.
In the present analysis we assume a constant magnetic field, 
since a possible field decay does not affect our results, at least
qualitatively.

\item

The torque due to the interaction between the magnetic moments of the
NS and of the secondary star. 
The torque acting on the NS due to this interaction is
\be
\label{e-mu}
\nn_{\rm dd} = \muns \times \bb_2(1)
\ee
(see e.g. \citealp{Jac75}), where  $\bb_2(1)$ is the RD
magnetic field in correspondence of the NS: this can
be written  in terms of the dwarf's magnetic moment 
$\murd$ as
\be
\bb_2(1) = \frac{3\, (\murd\cdot \n)\:\n - \murd}{a^3},
\ee
\citep{Jac75}, where $\n$ 
is the unit vector pointing from the RD to the NS.
We fix a non--rotating frame of reference centred on the NS, whose
$xy$~plane coincides with the orbital plane.  The torque~(\ref{e-mu})
also works on the orbital angular momentum
to make $\muns$, $\murd$ and $\n$ co-planar \citep{Kin90}. 
We therefore
assume that these vectors all lie in the orbital plane $xy$, so
\be
\n = - \left[
             \begin{array}{c}
               \cos\vartheta \\
               \sin\vartheta \\
               0
             \end{array} 
       \right]
\qquad
\muns/\mu_1 = \left[
             \begin{array}{c}
               \cos\phi_1\\
               \sin\phi_1\\
               0
             \end{array} 
       \right]
\qquad
\murd/\mu_2 = \left[
             \begin{array}{c}
               \cos\phi_2 \\
               \sin\phi_2 \\
               0
             \end{array} 
       \right],
\ee
being $\vartheta$ the orbital  true anomaly, $\phi_1$ and $\phi_2$ the 
inclinations
of the NS's and RD's magnetic moments with respect to the $x$-axis. 
After some algebra,  the torque~(\ref{e-mu}) reads
\be
\label{e-dd}
\nn_{\rm dd} = - \frac{\mu_1\: \mu_2}{a^3} \;
\biggl[ 2 \cos(\phi_2 - \vartheta)\: \sin(\phi_1-\vartheta) +
 \sin(\phi_2 - \vartheta)\: \cos(\phi_1-\vartheta)
\biggr]  \; \kk.
\ee
As a final remark concerning $\nn_{\rm dd}$,  we observe that it does
not contribute to lock the {\em period} of the NS to the orbital
value, but rather  its {\em phase angle} $\phi_1$
(see also the discussion in Appendix~\ref{s-equilibrium}). 
{\em Phase locking} is 
a stronger condition than  simple {\em period locking}. 
In the first case, the spin period and the orbital period coincide, but
$\phi_1$ may have any  value.
In the second case the orientation of $\phi_1$ is fixed by the dynamics.

Omitting the geometrical factor, we define the
characteristic time scale of the dipole-dipole torque~(\ref{e-dd})
\be
\label{e-time-dd}
t_{\rm dd} = \frac{\omega_1\:I_1}{|\nn_{\rm dd}|} \simeq
7 \times 10^{6} \;
\left(\frac{I_1}{10^{45}\g\cm^2}\right)\;
\left(\frac{P_1}{10\ms}\right)^{-1} \;
\left(\frac{\mu_1\;\mu_2}{10^{67}\G^2\cm^6}\right)^{-1}\;
\left(\frac{\porb}{6.67\hr}\right)^{2}\;\yr
\ee
This torque is weak when the NS rotates fast, 
but when its period is in the range of the hour, the time 
scale~(\ref{e-time-dd}) drops to few years, and is therefore important 
for the final equilibrium.

\item

A dissipative torque acting to synchronise the
NS spin with the orbital period. The nature and strength 
of this torque  have been investigated  by a host of authors, but there
is still no general consensus about it.
In their pioneering paper, \citet{Jos79} considered that the asynchronous
orbital revolution of the secondary in the primary's magnetosphere would
induce strong electrical currents in the secondary's atmosphere,
which are dissipated by Ohmic decay, 
leading the compact star to rotate synchronously with the orbit.
In a series of papers Campbell \citep{Cam83, Cam84a, Cam99, Cam05} 
considered  the role of turbulence
in the secondary star, and concluded that the turbulent dissipation would
be much more effective than  Ohmic decay in dissipating the currents
(but see \citealp{Lam85}).  Yet this torque becomes quite inefficient if 
$\omega_1\gtrsim\Omega_{\rm orb}$
($\Omega_{\rm orb}=2\,\pi/P_{\rm orb}$ is the orbital frequency), 
and therefore the compact star's initial spin period 
must not be too different from the orbital period to reach synchronism. 
Other authors suggest different mechanisms for the dissipative 
torque such as  unipolar induction \citep{Cha83, Wu02} 
or MHD torques \citep{Lam83, Kab86}.
Despite their  great differences,  essentially all the models 
write the dissipative torque as
\be
\label{e-disspar}
N_{\rm diss} = f \; \frac{\mu_1\; \mu_2}{a^3},
\ee
with a typical time scale
\be
t_{\rm diss}=(\omega_1-\Omega_{\rm orb}) \:I_1 /N_{\rm diss}
\ee
necessary to synchronise the primary; here
$f\simeq 1$ is a model dependent dimensionless factor.
The estimates of  $t_{\rm diss}$ given by the different models disagree by orders of magnitude: typical values 
range from about $10^3\s$ \citep{Kab86, Cha83, Wu02}, 
to more than $10^6\yr$ \citep{Cam83}
\footnote{
These times are meant as scaled from a white dwarf to a magnetar.}.
In view of the large uncertainty about $N_{\rm diss}$,  we  have chosen to
parametrise it. We would like to single out the dependence of this torque
on the asynchronism between the NS's rotation and the orbital period, so 
we write
\be
\label{e-diss}
\nn_{\rm diss} = 
- \; I_1 \;  \frac{\omega_1 - \Omega_{\rm orb}}{ \tau_{\rm syn}} \; \; 
\kk ,
\ee
where  the synchronisation time scale $\tau_{\rm syn}$ is a free parameter 
of our model. We will calculate the evolution of the NS's period for
several values of $\tau_{\rm syn}$. 
An important {\em caveat} is in order here. The  time 
$t_{\rm diss}$ strongly depends on the orbital separation $a$
(see Equation~(\ref{e-disspar})), 
so the dissipation torque is inefficient in  wide binaries.
Essentially all the models for Polar CVs call for the condition
$R_A\gtrsim a$ for synchronism.
As the strong dependency on $a$ is not explicit in the parameter 
$\tau_{\rm syn}$, we must check for consistency that in our synchronous
solutions the condition $R_A>a$ is fulfilled.

If the dissipation torque is inefficient (i.e. $\tau_{\rm syn}$
exceeds the age of RCW~103),  the magnetic torques are unable to 
synchronise the NS, contrary to what happens  in a Polar CV. 
In this case, 1E would be more similar to a IP system, governed by material torques, rather than 
a Polar-like system ruled by purely magnetic interactions.

\item
The accretion and propeller torques.
In its youth  the NS spins too  fast to accrete matter: 
this is the so-called  ``propeller'' regime (see e.g. \citealp{Rom03}). 
The mass coming  from
the companion star flies around the NS's magnetosphere,
extracts angular momentum and spins the NS down.
The exact form of the torque in the propeller stage is not well understood:
we adopt the formula  suggested by \citet[p.159]{Lip92} 
\be
\label{e-propeller}
\nn_p = - k_t \; \frac{\mu_1^2}{R_{\rm co}^3} \; \kk
\ee
where  $k_t$ is a constant to be determined later.

This torque is very efficient, as shown by its short 
characteristic  time scale
\be
\label{e-time-propeller}
t_{\rm p} =\frac{\omega_1\:I_1}{|\nn_p|} \simeq
3\times 10^2\; \; k_t^{-1} \;
\left(\frac{I_1}{10^{45}\g\cm^2}\right) \;
\left(\frac{\mu_1}{10^{33}\G\cm^3}\right)^{-2} \;
\left(\frac{M_1}{1.4\msun}\right) \;
\left(\frac{P_1}{10\ms}\right)\;\s.
\ee

The accretion torque  depends on the mass accretion rate $\dot{M}_1$ 
on the primary,  as well as on the arm's 
length  of the material stress.
In the case of a Polar,
the flow from the secondary is directly accreted as soon as it crosses 
the inner lagrangian point $L_1$: if the orbit is only moderately eccentric,
we may adopt the formula \citep{Pla64}
\be
\label{e-b1}
b_1 = a \; (0.500 - 0.227\: \log_{10} q) \simeq 9.5 \times 10^{10} \cm
\ee
to approximate the distance $b_1$ between the inner Lagrangian point 
$L_1$ and  the  NS centre. The  torque's arm $b_1$ yields 
\be
\label{e-accretion}
\nn_a = \dot{M}_1 \; (G\; M \;b_1)^{1/2} \;  \kk,
\ee
whose associated characteristic time scale is (up to factors of order unity)
\be
\label{e-time-accretion}
t_a \simeq
1.7\times 10^2 \;
\left(\frac{I_1}{10^{45}\g\cm^2}\right) \;
\left(\frac{\dot{M}_1}{10^{13}\g\s^{-1}}\right)^{-1}\;
\left(\frac{\porb}{6.67\hr}\right)^{1/3}
\left(\frac{P_1}{6.67\hr}\right)^{-1}\;\yr.
\ee
We describe both the accretion and the propeller regimes with a 
unique accretion/propeller torque
\be
\nn_{\rm ap} = \nn_p +  \nn_a. 
\ee
We write $\dot{M}_1$ as a function of the mass transfer
rate $\dot{M_2}$ from the secondary. We postulate the relation
\be
\label{e-m1m2}
\dot{M}_1 = \alpha \; |\dot{M}_2|.
\ee
where $\alpha\simeq 0$ if $N_a\ll N_p$ and $\alpha\simeq 1$ if
$N_a\gg N_p$. Since the accretion torque is dominant if $b_1\ll R_{\rm co}$
and the opposite inequality is true if the propeller is dominant, we choose
\be
\alpha = \frac{R_{\rm co}^3}{b_1^3 + R_{\rm co}^3},
\ee
which may be rewritten
\be
\alpha = \frac{1}{1 + \lambda \; (\omega_1/\Omega_{\rm orb})^2},
\ee
where
\be
\lambda = \lambda(q) \equiv (b_1/a)^3 \;  \; (1 + q)
\ee
is a function of the mass ratio $q$ only (cf. Equation~(\ref{e-b1})).

We determine the constant $k_t$ so that $N_{\rm ap}=0$ when 
$R_{\rm co}=b_1$, i.e.  
when the centrifugal barrier opens. After some algebra, we retrieve 
the formula
\be
\label{e-ap}
\nn_{\rm ap} =
|\dot{M}_2| \; (G\:M\:b_1)^{1/2}  \;
\left[ \frac{1}{1 + \lambda \; (\omega_1/\Omega_{\rm orb})^2}  - 
\frac{\lambda}{2} \; 
\left(\frac{\omega_1}{\Omega_{\rm orb}}\right)^2\right]
 \;  \kk
\ee
In the accretion limit $\omega_1\ll \lambda^{-1/2}\:\Omega_{\rm orb}$ 
the propeller torque is  small,
and this expression reduces to that of pure accretion~(\ref{e-accretion}).
On the other hand, in the propeller regime  
$\omega_1\gg \lambda^{-1/2}\Omega_{\rm orb}$
the first term in square brackets  is negligible, and the torque reduces to
pure propeller $N_{\rm ap}\propto -\omega_1^2$,  as in 
Equation~(\ref{e-propeller}).

Incidentally, we note that the  NS spin equilibrium period
$P_{\rm eq} = \lambda^{-1/2} P_{\rm orb}$
 defined by $N_{\rm ap}=0$ is the Keplerian period at $L_1$ of a mass 
 orbiting the NS, and
coincides with   that predicted by 
the diamagnetic blobs accretion model  put forward  by \cite{Kin93} 
and \citet{Wyn95} for highly magnetised  Cataclysmic Variables.

\end{enumerate}

The torques analysed in this Section are the building blocs for the
overall torque acting on the NS: this will be made explicit in
Section~\ref{s-polar}.

\subsection{Torques Acting on the Secondary Star}
\label{s-rd}

The reason why we consider also the torques on the secondary star is that they
may couple to the orbit to alter $P_{\rm orb}$, thus  affecting synchronism
(see also \citealp{Kin90} for a fuller discussion).
In general, the torques on the secondary star  are:

\begin{enumerate}

\item
The magnetic braking torque. If the RD is magnetic and  blows a wind, 
this is coupled to the magnetic field and may extract angular momentum
from the secondary. This torque only depends on the 
RD period and not on the orbital one, so  magnetic braking
does not directly affect the orbital period.
However, it may couple with the tidal torque (discussed immediately below) 
to transfer angular momentum from the orbit to the RD.

\item
The tidal torque reduces the relative  rotation of the RD
and the orbit, and of course vanishes if the secondary rotates synchronously.
This torque may also redistribute  angular momentum
between the orbit and the RD. 
Stars with convective envelopes (which is most likely to be the case of
the our companion star) in binary systems with orbital periods of few hours
have tidal synchronisation times~$\approx 10^2\yr$ (Equation~\ref{e-tides}), 
so we shall assume that the RD is always 
locked to the orbital period. 
The magnetic braking extracts angular momentum from the secondary,
which is then taken away from synchronism. The tidal torque, on the  other
hand, is very effective in locking the secondary back to orbital synchronism.
As a consequence, the  synergic action of these torques pumps angular momentum from  the orbit to the secondary, keeping it in synchronous rotation.
\citep{Ver81}.

If the secondary is not synchronous, there are two effects.
First, the dipole-dipole torque~(\ref{e-dd}) keeps oscillating from 
positive to negative, so it  cannot  lock the NS's phase.
Second,  the secondary can exchange some if its intrinsic angular momentum
with the orbit. In view of our estimates (Section~\ref{s-orb} below),
we do expect that the amount of transferred angular momentum is too
small to affect  the orbit.
Summarising, the secondary's possible  lack of synchronism does not directly
alter the primary's period locking, but may  invalidate its
phase locking.

\item

The dipole-dipole torque. This is exactly the same kind of 
torque~(\ref{e-dd})   acting 
on the primary: its expression is readily derived 
by swapping  $\phi_1$ and $\phi_2$ in  Equation~(\ref{e-dd}):
\be
\label{e-rd}
\nn'_{\rm dd} = - \frac{\mu_1\: \mu_2}{a^3} \;
\biggl[ 2\, \cos(\phi_1- \vartheta)\:  \sin(\phi_2-\vartheta)
+
\sin(\phi_1- \vartheta)\:  \cos(\phi_2-\vartheta) \biggr]  \; \kk
\ee
This torque depends on the true anomaly as well as on the stars' 
magnetic moments, and is able to exchange angular momentum between the 
secondary and the  orbit.
The synchronisation time scale for this torque is 
\be
\label{e-time-dd-rd}
t'_{\rm dd} = \frac{\omega_2\:I_2}{|\nn'_{\rm dd}|} \simeq
3 \times 10^{9} \;
\left(\frac{I_2}{10^{54}\g\cm^2}\right)\;
\left(\frac{P_2}{6.67\hr}\right)^{-1} \;
\left(\frac{\mu_1\;\mu_2}{10^{67}\G^2\cm^6}\right)^{-1}\;
\left(\frac{\porb}{6.67\hr}\right)^{2}\yr,
\ee
where $P_2$ is the period of secondary.
Since $t'_{\rm dd}$ is much longer than the age of RCW~103,
the RD phase is not likely to be locked,
although its period is most likely synchronous with the orbit.

\end{enumerate}

In summary, we assume that the secondary's rotation is synchronous  with the 
orbit. The instantaneous phase  $\phi_2$ (i.e. the position
of the RD magnetic moment ${\bf\mu}_2$) is 
\be
\phi_2 = \Omega_{\rm orb} \; t + \phi_2',
\ee
where the phase $\phi_2'$ is arbitrary due to the lack of phase locking.

\subsection{Orbital Torques}
\label{s-orb}

In this Section we derive and
discuss the torques acting on the orbit of the binary system.
Since some torques and  the mass transfer
alter the orbital period, they must be taken into account to
study the synchronisation. 
As usual, we suppose that the orbit is circular, so 
the orbital angular momentum reads
\be
\label{e-jorb}
\jj_{\rm orb} 
= \left(\frac{G \; a}{M}\right)^{1/2} \; M_1 \;M_2\;\kk
\ee
The rate of change of $\jj_{\rm orb}$ is due to the action of all the
stellar torques containing the true anomaly $\vartheta$ 
(or the orbital frequency $\Omega_{\rm orb}$),
plus the tidal/magnetic braking torque $\nn_{\rm mag}$ due to the 
angular momentum exchange with the secondary discussed in Section~\ref{s-rd}:
\be
\label{e-jdot}
\dot{\jj}_{\rm orb} = -\nn_{\rm dd} - \nn'_{\rm dd} -
  \nn_{\rm ap} - \nn_{\rm diss}  - \nn_{\rm mag}.
\ee
Gravitational radiation has not been included: it is generally important for systems with short period ($P_{\rm orb} \lesssim 2\hr$), and it is immaterial here. 

The torque $N_{\rm mag}$  is \citep{Ver81}
\be
N_{\rm mag}
\simeq
0.5 \times 10^{-28} \; I_2\; R_2^2 \; \Omega^3_{\rm orb} 
\ee
(in cgs units). This torque drags orbital angular momentum on the
time scale
\be
t_{\rm mag} = J_{\rm orb}/N_{\rm mag}
\simeq
10^8 \; \left(\frac{m}{0.3\msun}\right) \; 
\left(\frac{P_{\rm orb}}{6.67\hr}\right)^{10/3} \; 
\left(\frac{I_2}{10^{54}\g\cm^2}\right)^{-1} \; 
\left(\frac{R_2}{3.3\times 10^{10}\cm}\right)^{-2}\yr, 
\ee
where $m$ is the reduced mass of the binary system. Since this time scale is 
very long, the orbital angular momentum yield to the secondary may be
neglected.

Ignoring the exchanges of orbital angular momentum with the RD,
we focus on the exchanges with the rapidly spinning NS.
Adopting the parametrisation~(\ref{e-m1m2}) for the mass accretion rate,
the  logarithmic derivative of the modulus of Equation~(\ref{e-jorb}) gives
\be
\frac{\dot{J}_{\rm orb}}{\jorb} = \onehalf\: \frac{\dot a}{a} -
\frac{|\dot{M}_2|}{M_2} \; \frac{2\, (1-\alpha q^2) + q\, (1 -\alpha)}{2\,(1+q)}
\ee
Plugging the expressions of the relevant torques into~(\ref{e-jdot}),
and using the Kepler relation $a\propto\Omega_{\rm orb}^{-2/3}$ we retrieve our
equation for the evolution of the orbital frequency
\begin{eqnarray}
\label{e-orbit}
\nonumber
-\frac{\dot{\Omega}_{\rm orb}}{\Omega_{\rm orb}} &= 
\frac{3\;I_1}{m\; a^2\; \Omega_{\rm orb}} \; 
\biggl[3 \: \frac{\mu_1\: \mu_2}{I_1\;a^3} \;
 \sin(\phi_1+ \phi_2-2\,\vartheta) +
\frac{\omega_1 - \Omega_{\rm orb}}{\tau_{\rm syn}} -
\\
\nonumber
&-\frac{|\dot{M}_2| \; (G\:M\;b_1)^{1/2}}{I_1} \; 
\left[\frac{1}{1+\lambda\:(\omega_1/\Omega_{\rm orb})^2} 
- \onehalf\lambda\, \left(\omega_1/\Omega_{\rm orb}\right)^2 \right]
\biggr] +
\\
&+\frac{3}{2} \; \frac{|\dot{M}_2|}{M_2} \; 
\frac{2\, (1-\alpha\,  q^2) + q\, (1 -\alpha)}{1+q}.
\end{eqnarray}
We briefly discuss the magnitude of the terms on the right-hand side
of Equation~(\ref{e-orbit}).
The terms in square brackets are the same dipole-dipole, dissipative and
accretion/propeller torques (with allowance of signs and  geometrical
factors) acting on the NS; all these terms are multiplied by the 
small factor $\zeta=3\; I_1/m\, a^2\,\Omega_{\rm orb}\simeq 10^{-6}$, 
showing that
the time scales for the orbital evolution are longer than the time scales
for the NS' spin evolution by a factor $\zeta^{-1}\simeq 10^6$.

Mass transfer unbalances the relative masses of the two
components,  and only becomes significant over the long time scale
\be
t_2 \simeq M_2/|\dot{M}_2| =
2.5\times 10^{12}\;
\left(\frac{M_2}{0.4\msun}\right) \;
\left(\frac{|\dot{M}_2|}{10^{13}\g\s^{-1}}\right)^{-1}\yr.
\ee

In summary, the orbital torques are characterised by
time scales much longer than the current age of RCW~103; 
the orbit may be then be regarded as fixed, and in the following
we only consider the torques on the NS.
The results presented in the next Section are based on this hypothesis;
however we had an additional check of its validity by running a numerical 
calculation where the orbital parameters were allowed to evolve 
according to Equation~(\ref{e-orbit}).

\section{Evolution of the NS Rotation}
\label{s-polar}

We adopt $\phi_1$ (the phase of the NS magnetic moment) 
as the NS Euler free rotation 
angle  \citep[e.g][]{Lan69}. Putting all the 
torques~(\ref{e-md}),~(\ref{e-dd}),~(\ref{e-diss}) and~(\ref{e-ap}) 
together, 
the equation of motion for the NS's phase angle $\phi_1$ reads
\begin{eqnarray}
\label{e-spin}
\nonumber
&\dot{\phi}_1 = \omega_1
\\
&I_1 \; \dot{\omega}_1 = 
- \frac{\mu_1\: \mu_2}{a^3} \;
\biggl[ 
2\, \cos(\phi_2 - \vartheta)\: 
\sin(\phi_1 - \vartheta)
+ \sin(\phi_2 - \vartheta)\: 
\cos(\phi_1 - \vartheta)\biggr] - \twothirds \:
 \frac{\mu_1^2}{c^3} \;  \omega_1^3 -
\nonumber
\\
&- I_1 \: \frac{\omega_1 - \Omega_{\rm orb}}{\tau_{\rm syn}} +
|\dot{M}_2| \; (G\:M\; b_1)^{1/2} \;  
\left[ \frac{1}{1 + \lambda \: (\omega_1/\Omega_{\rm orb})^2} - 
\frac{\lambda}{2}  \left(\omega_1/\Omega_{\rm orb}\right)^2  \right]
\end{eqnarray}
where  the true anomaly $\vartheta = \Omega_{\rm orb} \, t$
is assumed to be zero at $t=0$. As noticed above, we keep  the
orbital parameters constant, as their variation due to the angular momentum exchanged with the  stars is negligible.

As shown in Appendix~\ref{s-equilibrium}, Equation~(\ref{e-spin}) 
admits an  asymptotic constant solution, i.e. a stable NS equilibrium
spin period.  
In the absence of material torques  ($\dot{M}_2 =0$), this period 
is attained thanks to the second and the third term in Equation~(\ref{e-spin}),  leading to  full synchronism with the orbit. 
The magneto-dipole torque brakes the initial fast NS rotation, and 
later on the dissipative torque brings the NS spin to orbital 
synchronism.  In presence of  material torques, it 
is the interplay between propeller and accretion torques that drives 
the young magnetar into equilibrium  (as implied by Equation~(\ref{e-peq})).   
In this case, the orbital period does not  coincide with the NS's spin
equilibrium $P_{\rm eq}$, 
but it is quite close to it ($P_{\rm eq}\simeq 0.5-0.7~P_{\rm orb}$).
Even in this case the NS magnetosphere keeps filling the binary Roche lobe,
warranting the consistency of our results.

We integrated Equations~(\ref{e-spin}) with a step-adaptive fourth
order Rosenbrock algorithm \citep{Pre02}.  This choice is motivated by the possible
stiffness of the problem due the diverse time scales of the torques.  
The NS and RD magnetic moments are the free parameters 
of the model, together with $|\dot{M}_2|$  and $\tau_{\rm syn}$. 
The magnetic moments of the NS and the RD are
$\mu_1=10^{33} \G\cm^3$ and $\mu_2=10^{34} \G\cm^3$
respectively. The NS's moment of inertia has the standard value of
$I_1= 10^{45} \g\cm^2$. We let the mass transfer rate and 
the synchronisation time to vary in each run, in order to explore different 
evolution paths.
The initial phases $\phi_1(t=0)$ and $\phi_2(t=0)$ have been randomly
chosen, with an initial NS rotation period of $3 \ms$. The RD has
been always considered synchronous.  In all our runs the NS final
equilibrium spin period has been  set equal to $6.67\hr$ 
(see Appendix~\ref{s-equilibrium}), and we have derived the orbital
period in a self-consistent manner.

For consistency, we have also checked that the condition that $R_A\gtrsim a$ 
is always met in our runs.
The NS Roche lobe is dominated by the star's magnetic field, and
the mass from the secondary is  effectively channelled from  $L_1$ to the 
star's polar caps without forming an accretion disc, consistent with 
the torque~(\ref{e-ap}) used in our calculations.

\begin{figure}
\begin{center}
\includegraphics[width =90mm, angle = 90]{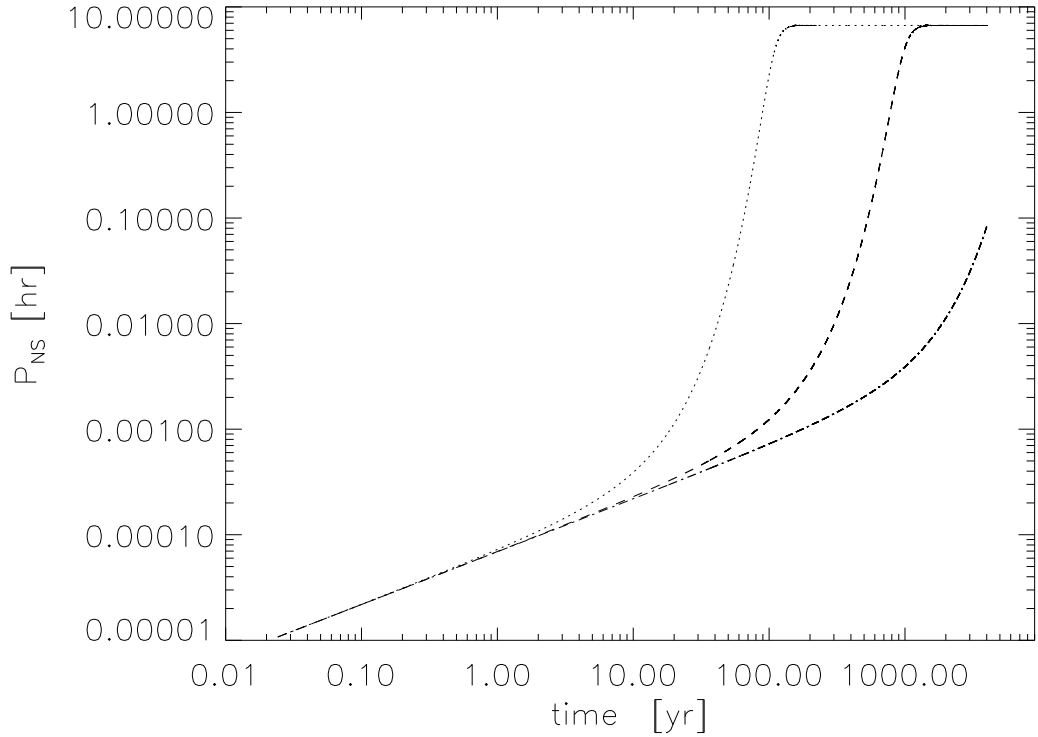} 
\end{center}
\caption{Evolution of the NS's rotation period without  
mass transfer ($\dot M_2=0$). The dotted line refers to the synchronisation
time  $\tau_{\rm syn}=10\yr$, the dashed line to  $\tau_{\rm syn}=100\yr$
and the dot-dashed line to $\tau_{\rm syn}=1000\yr$.
In all cases the orbital period is $P_{\rm orb}=6.67\hr$.
}
\label{f-mdot0E00}
\end{figure}

It is convenient to present first our result in the case where the mass
transfer  rate $\dot{M}_2$ vanishes. The material (i.e. accretion 
and propeller)  torques are always zero, and 
the evolution is controlled uniquely  by the magnetic torques.
Figure~\ref{f-mdot0E00} plots the evolution 
of the NS' period  for different values of the 
synchronisation time: $\tau_{\rm syn}=10\yr$, 
 $\tau_{\rm syn}=100\yr$ and
 $\tau_{\rm syn}=1000\yr$, respectively.

The dissipative magnetic torque plays, in this case, a key role 
in driving the NS spin into orbital synchronisation, 
attained within $\sim 2000\yr$ 
if $\tau_{\rm syn}$ is not longer than  
few hundred years. A magnetar-like field and a close orbit 
are necessary and this is consistent with
estimates given in the literature, possibly 
with the exception of
\citet{Cam83,Cam05} (see also the discussion in Section~\ref{s-ns}).

\begin{figure}
\begin{center}
\includegraphics[width =90mm, angle = 90]{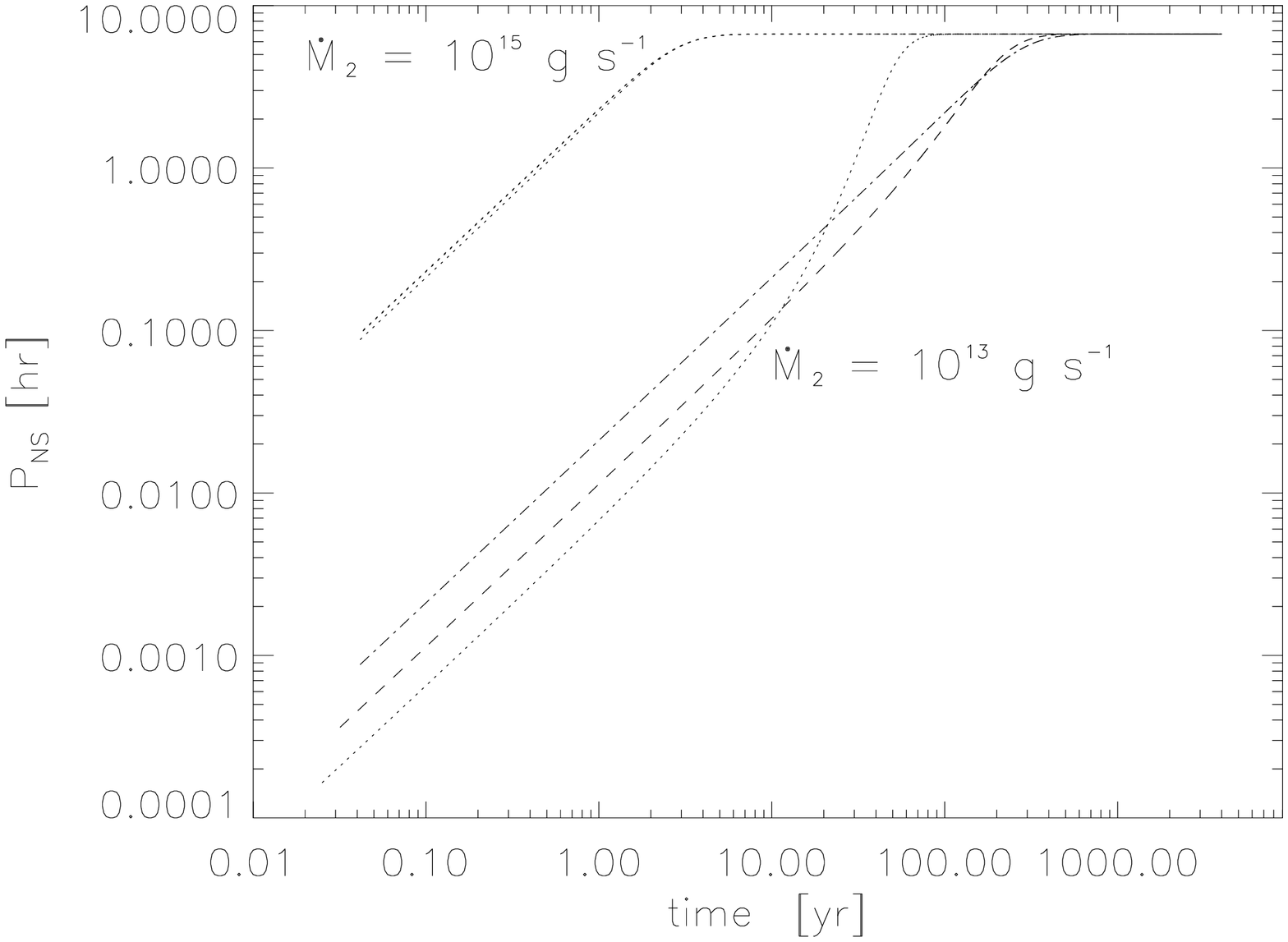} 
\vskip 10mm
\includegraphics[width =90mm, angle = 90]{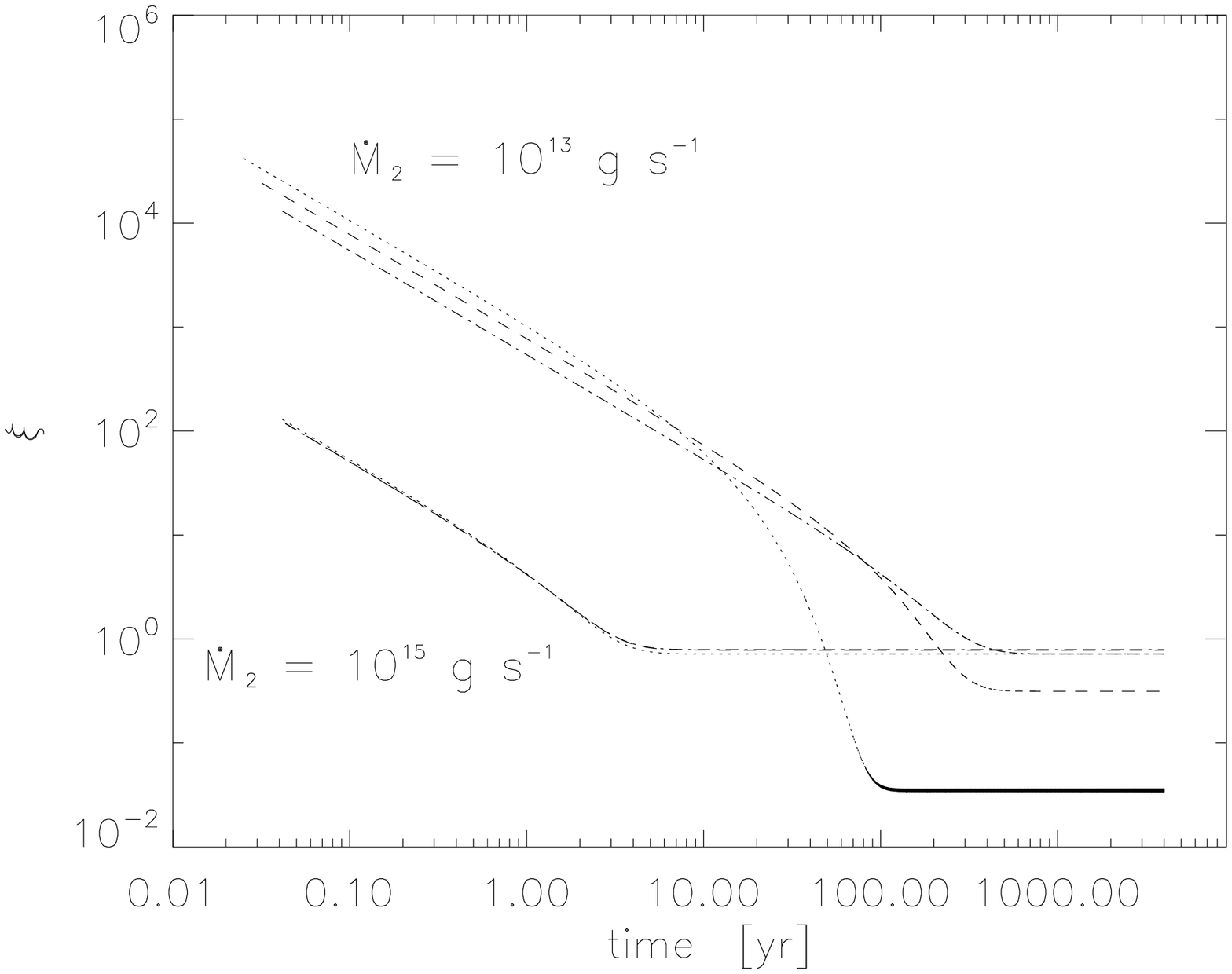} 
\end{center}
\caption{Evolution of the NS's rotation period (upper panel)
and of the normalised synodic frequency 
$\xi=(\omega_1-\Omega_{\rm orb})/\Omega_{\rm orb}$ (lower 
panel). The two groups of lines refer to the mass transfer rates 
$|\dot{M}_2|=10^{13}\g\s^{-1}$ and $|\dot{M}_2|=10^{15}\g\s^{-1}$.
 The lines' coding is the
same as in Figure~\ref{f-mdot0E00}.
For $|\dot{M}_2|=10^{13}\g\s^{-1}$ the orbital periods are 
$P_{\rm orb}=6.9\hr$ (for $\tau_{\rm syn}=10\yr$),
$P_{\rm orb}=8.7\hr$ (for $\tau_{\rm syn}=100\yr$) and
$P_{\rm orb}=11.5\hr$ (for $\tau_{\rm syn}=1000\yr$).
For $|\dot{M}_2|=10^{15}\g\s^{-1}$  the orbital periods are
$P_{\rm orb}=11.4\hr$  ($\tau_{\rm syn}=10\yr$), and
 $P_{\rm orb}=11.9\hr$ (both for $\tau_{\rm syn}=100\yr$ and 
$\tau_{\rm syn}=1000\yr$).
Note that for the mass transfer rate  $|\dot{M}_2|=10^{15}\g\s^{-1}$  
the condition $R_A>b_1$ is violated (see text for more details).
}
\label{f-mdot1315}
\end{figure}

The situation changes if the mass transfer rate is non-negligible.
Figure~\ref{f-mdot1315} plots the evolution of $P_1$ and of the 
synodic normalised frequency 
\be
\xi=(\omega_1-\Omega_{\rm orb})/\Omega_{\rm orb}, 
\ee
for $|\dot{M}_2|=10^{13}\g\s^{-1}$. 
Here the material torques are
relevant, and   lead to the equilibrium spin period 
within few hundred years, fairly independent on the exact 
value of $\tau_{\rm syn}$.
The  orbital period is longer than the NS's equilibrium spin period;
we find 
$P_{\rm orb}=6.9\hr$ if  $\tau_{\rm syn}=10\yr$, 
$P_{\rm orb}=8.7\hr$ if  $\tau_{\rm syn}=100\yr$ and
$P_{\rm orb}=11.5\hr$ for $\tau_{\rm syn}=1000\yr$.
An additional run with  $\tau_{\rm syn}=10^7\yr$ shows that
the magnetar attains its spin equilibrium within~$\sim 10^3\yr$, 
with an  orbital period of~$\sim 12\hr$, so that in all cases 
we find  $P_{\rm spin}/P_{\rm orb}\gtrsim 0.5$. In magnetic CVs 
period  ratios larger than~$0.5$ correspond to stream-fed accretion onto
the white dwarf~\citep{Nor04}.
Consistently, we find  also here that  the condition $R_A>b_1$ 
is always  met:  the NS's Roche lobe is filled by the 
magnetosphere, and no accretion disc can form.
In addition, for $\tau_{\rm syn} < 1000\yr$, the stronger condition
$R_A>a$ is also met: this warrants that this (parametric) synchronisation time 
is consistent with the requirement of a strong magnetic interaction between the
stars (Section~\ref{s-ns}).

In the same  Figure~\ref{f-mdot1315} we show the evolution  of 
$P_1$ and $\xi$ for the much larger  mass transfer rate
$|\dot{M}_2|=10^{15}\g\s^{-1}$, corresponding to the peak 
X-ray luminosity observed in 1E.
This value should provide an upper limit to 
the influence of the mass accretion rate on the period evolution.
For this  value of $|\dot{M}_2|$,  the condition $R_A>b_1$ 
is violated, but since $R_A$ is still larger than the circularisation 
radius $R_{\rm circ}$ of the accreting flow, the transferred mass  is  
unable to circularise and form a disc \citep[see e.g.][]{Fra02}. 
For such a strong material torque, the equilibrium period is 
attained in 
less than~$~\sim 10\yr$ during  the early ``propeller'' phase which 
extracts a great
amount of angular momentum from the spinning young magnetar.

\section{Discussion}
\label{s-discussion}

In the previous Section we have shown that if 1E hosts a highly
magnetic NS orbiting a RD, the NS is spun down rapidly to an
asymptotic equilibrium period, either by magnetic and/or material
torques. The magnetar-like NS loses memory of the spin acquired at
birth, and approaches a spin period close to the binary orbital
period.

If the material/propeller torques have been small 
for an appreciable fraction
of the system's age, then magnetic torques have been dominating since 
birth,  and    orbital synchronism is attained within $2000\yr$ if 
$\tau_{\rm syn}\lesssim 100\yr$.
The X-ray emission in this case should be dominated by mechanisms similar
to those seen in Magnetars (Section~\ref{s-setup}).

On the other hand, if the material torques have been important,
1E is more similar to a young accreting low-mass binary. 
The close resemblance of
the light curve of 1E with Polars and Intermediate
Polar Cataclysmic Variables  suggests that mass transfer 
currently takes place
in 1E via filling of the RD's Roche lobe. 
(Note that this does not exclude the action of magnetic torques in
the earlier life of the source).
Although the magnetic torques are unable to synchronise the system,
the  hypothesis of a magnetar-like field is still essential here,  
since the NS magnetosphere must dominate the star's
Roche lobe. If $B_1\simeq 10^{15}\G$ and the orbital period is of
$6-12\hr$, then $R_A>b_1$ and the flow
from the RD provides a material torque with arm of length~$\sim b_1$,
which brings the NS to the observed equilibrium  spin period  
$P_{\rm eq}=6.67\hr$  within~$\sim 100\yr$, 
independent of the magnetic torques. 
The orbital period in this case still depends on $\tau_{\rm syn}$, 
and is longer than  $P_{\rm eq}$. In the limit of a 
very inefficient dissipative torque we find  $P_{\rm orb}\sim 2 P_{\rm eq}$ 
(i.e. $P_{\rm orb}\simeq 12\hr$)
\footnote{
It is worth spending here few words on the upper limits of the orbital period. 
If the companion star tightly follows the mass-radius relation valid
for low main sequence stars, then the orbital period is unlikely to
exceed by much  $P_{\rm orb}\sim 12\hr$, since this  binary separation
brings the NS to its spin equilibrium period 
due to the sole action of the material torques. 
These imply a Roche-lobe filling donor star, which in our calculations
we assumed to be of $0.4\msun$, close to the upper limit set by the 
optical/IR observations. Were the system wider than this, the star
should fall within its Roche lobe, making mass transfer impossible, as
well as the action of the material torques.
On the other hand,  mass transfer is possible if the companion star is larger than a main sequence star of the same size, e.g. because it absorbed energy 
from the early relativistic wind emitted by the NS.
In this case the limits on the orbit's size are more uncertain,
but  by analogy with the period distribution of {\sl bona fide} Polars and IPs, 
we may estimate that the orbital period is unlikely to exceed by much 
$P_{\rm orb}\sim 12\hr$.
The current observations, on the other hand, are much less constraining.
Our X-ray data do not yield any evidence for 
a periodical modulation larger than $6.67\hr$.
In any case, it would be quite  difficult to detect a modulation
at a few percent level and 
a long periodicity ($10-15\hr$) superimposed 
on the very strong pulsation at $6.67\hr$.
}.

A remark about our use of the circular approximation.
As can be seen from Equation~(\ref{e-spin}), which rules the secular
evolution of the NS's spin, the circular approximation enters:
i)~the distance between the components in the dipole-dipole interaction
;
ii)~the orbital frequency $\Omega_{\rm orb}$ in the dissipation term
and the accretion/propeller torque 
;
iii)~the arm $b_1$ (the distance between the neutron star and the inner 
Lagrangian point), which is calculated in the Roche formalism.
Note that  (whatever the eccentricity) 
point~iii) is immaterial if the RD fits within its Roche lobe.
In this case, there is no mass transfer, and the accretion/propeller 
torque vanishes; if the companion is large enough to allow mass accretion, 
the arm $b_1$ only enters as a weak power in the torque, and a 
possible deviation from the circular geometry is unlikely to affect 
this  torque to a significant amount.
As far as points~i) and~ii) are concerned, the introduction of 
a non-zero eccentricity would induce an orbital modulation in the 
dipole-dipole and in the dissipation torque. This 
modulation is not able to alter the qualitative results presented in this
paper. 

In this model,  1E  would be
the first young low mass binary observed in X-rays in the first 
$\sim 2000 \yr$
of its evolution, and there might be a connection between the
magnetar nature of the NS and this premature onset of an X-ray phase
if it comes from accretion. 
Two questions arise: 
\begin{enumerate}
\item 
Does the pulsar wind
of the young magnetar induce a mass loss from the RD? 
Can this wind vaporise the RD? How does this effect constrains
the initial secondary's mass, or the NS initial period?
\item 
Looking ahead,  what is the 
subsequent evolution of 1E? 
\end{enumerate}

As far as the first question is concerned, 
the strong wind of photons and
relativistic particles of the magnetar may perturb the  
RD thermal equilibrium, and may in principle cause its partial evaporation.
The problem we address here differs from the evaporation of a 
main sequence star by a millisecond 
pulsar in a narrow binary system
(e.g. \citealp{Rud89a,Rud89b, Tav92}).
In these systems, the NS magnetic field is low ($10^8-10^9\G$) 
and its weak irradiation lasts for more than~$10^8\yr$.
In a millisecond pulsar environment,
the main sequence star has enough time to adjust to thermal equilibrium
during the evaporation. In the present case this assumption is most likely 
invalid, since the irradiation from the magnetar is much more intense
and short-lived (as shown below).

\bigskip

If the magnetar in 1E has a birth period $P_0=2\,\pi/\omega_0$ 
of few milliseconds, it holds a rotational energy $E_{\rm rot}\simeq 2\times 10^{51}(P_0/3~\rm{ms})^{-2}\erg$, which  is dissipated mainly through magnetic dipole losses in the early years,
when the propeller torques are still ineffective.
The RD has a binding energy $E_{\rm RD}\simeq G\,M_2^2/R_2\simeq 10^{48}
(M_2/0.4\msun)^2\erg,$ so 
the RD needs to absorb only  $\sim 0.1\%E_{\rm rot}$ to be evaporated.

We can estimate the mass loss rate from the RD by comparing the NS
energy loss  (and impinging on the RD)  with the 
kinetic energy necessary to sustain a wind, regardless of
the physical interaction process
\footnote{
The wind of particles from an ordinary millisecond pulsar 
is composed by energetic photons (X-rays or $\gamma$-rays) 
and~$\rm TeV$  $\rm e^\pm$ pairs \citep{Rud89a}.
In the case of a magnetar, the occurrence of high energy pairs is
debatable, since the strong magnetic field may prevent their formation
due to photon splitting \citep{Dun92}. With this caveat in mind,
we estimate how deep the impinging wind penetrates the companion star.
Photons are stopped in the stellar atmosphere, having a mean free path
of few $\cm$  in a gas with density~$\sim 1 \g\cm^{-3}$.
The electrons impinging on the companion star yield their energy chiefly 
for Inverse Compton losses on the radiation field,  and  for relativistic 
Bremsstrahlung. The  Inverse Compton and Bremsstrahlung stopping lengths are
$l_{\rm IC} \simeq 10^6\cm$  and 
$l_{\rm BS} \simeq  10^5 \cm$, respectively 
(\citealp{Lan80} p. 467):
since $l_{\rm BS}$ and $l_{\rm IC}$ are both small in comparison to
the stellar radius,  essentially all the energy carried by the
relativistic wind is deposited  in the star's  envelope.
}.
In more detail,  the spin-down luminosity of the magnatar decays with time
as 
\be
\label{e-sdl}
L_{\rm SD}(t) = 
\frac{1}{2} \frac{I_1 \;\omega_0^2}{\tau_{\rm md}} \;
\left(1 + \frac{t}{\tau_{\rm md}}\right)^{-2},
\ee
where $\tau_{\rm md}$ is the magneto-dipole decay time
\be
\label{e-tmd}
\tau_{\rm md} = \frac{3\; c^3 I_1}{4\; \mu_1^2 \; \omega_0^2}
\simeq 
4.6\times 10^3 \s \;  
\left(\frac{\mu_1}{10^{33}\G\cm^3}\right)^{-2} \; 
\left(\frac{P_0}{3\ms}\right)^2.
\ee
Equations~(\ref{e-sdl}) and~(\ref{e-tmd}) show that
(i) the magnetar is braked to a period of $1\sec$
after~$\sim16\yr$; 
(ii) most of the rotational energy is dissipated in the first year;
(iii) the spin-down luminosity decays by a factor~$\sim 10^{10}$ 
over this very short time.
At $P\sim10\sec$ the magnetar in 1E brakes under the action of the magnetic 
and material torques,  but at this period the bulk of its rotational energy has already been dissipated, and is  but a tiny fraction of the companion's binding energy. For this reason, any later mechanism that brakes the NS and which
may --~in principle~-- transfer energy from  the 
NS to the companion,  is most likely to have little effect on the latter.

We suppose that  this relativistic wind is emitted isotropically, so a
fraction $(R_2/2\,a)^2$ hits the 
companion star; a fraction $\eta$ of this  impinging power
extracts an evaporation wind from the RD.
Following  \citet{VdH88}, we equate the effective power of the 
relativistic wind with the evaporation wind to estimated $\dot{M}_2$:
\be
\label{e-wind1}
\eta\; (R_2/2 a)^2 \; L_{\rm SD}=
- \frac{G\: M_2}{R_2}\; \frac{d M_2}{d t},
\ee
where we have assumed that the evaporation wind speed is close to
the escape velocity from the star's atmosphere.
We obtain  the differential  equation
\be
\label{e-evaporate}
- \frac{G \: M_2}{R_2} \; \frac{d  M_2}{d t} 
=
\frac{1}{2} \; 
\eta \; \left(\frac{R_2}{2 \, a}\right)^2 \;
 \; \frac{I_1 \;\omega_0^2}{\tau_{\rm md}} \;
\left(1 + \frac{t}{\tau_{\rm md}}\right)^{-2}
\ee
for the evolution of the star's mass $M_2(t)$, which
can be solved once we have
specified the relation between $M_2$ and the star's radius $R_2$.
This relation is uncertain, so we consider two limiting cases aiming 
to bracket the actual behaviour.
\begin{itemize}
\item
On the  one hand, we assume  that $R_2$ is given, instant by instant, 
by the relation~(\ref{e-lms})
valid for lower main sequence stars. This is the most optimistic
limit, since it assumes that the companion star has  enough
time to reduce its size as a  reaction to the evaporating flux. This may be 
not very realistic, however.  The star reacts to the relativistic wind on 
the  Kelvin-Helmholtz time for the envelope involved in the
energy deposition
\be
\label{e-kh}
\tau_{\rm KH} \simeq \frac{G \: M_2^2}{R_2\: L_2} \; \frac{l_{\rm BS}}{R_2}
\simeq 10^4 \yr,
\ee
which is much longer than the magneto-dipole time scale
(\ref{e-tmd}); $L_2$ is the RD's luminosity, and $l_{\rm BS}$ is the 
penetration depth of the relativistic wind, where thermal equilibrium is
perturbed  (see note).

\item
On the other hand, we may suppose that 
the star's envelope swells up,  filling  its own Roche lobe. In this case, 
we assume for $R_2$ the size of the Roche lobe at the evaporation onset. 
This is the most pessimistic limit, since in this case 
the section presented by the star to the impinging flux is the largest, 
resulting  in a fast evaporation rate.
\end{itemize}

\begin{figure}
\begin{center}
\includegraphics[width = 90mm, angle = 90]{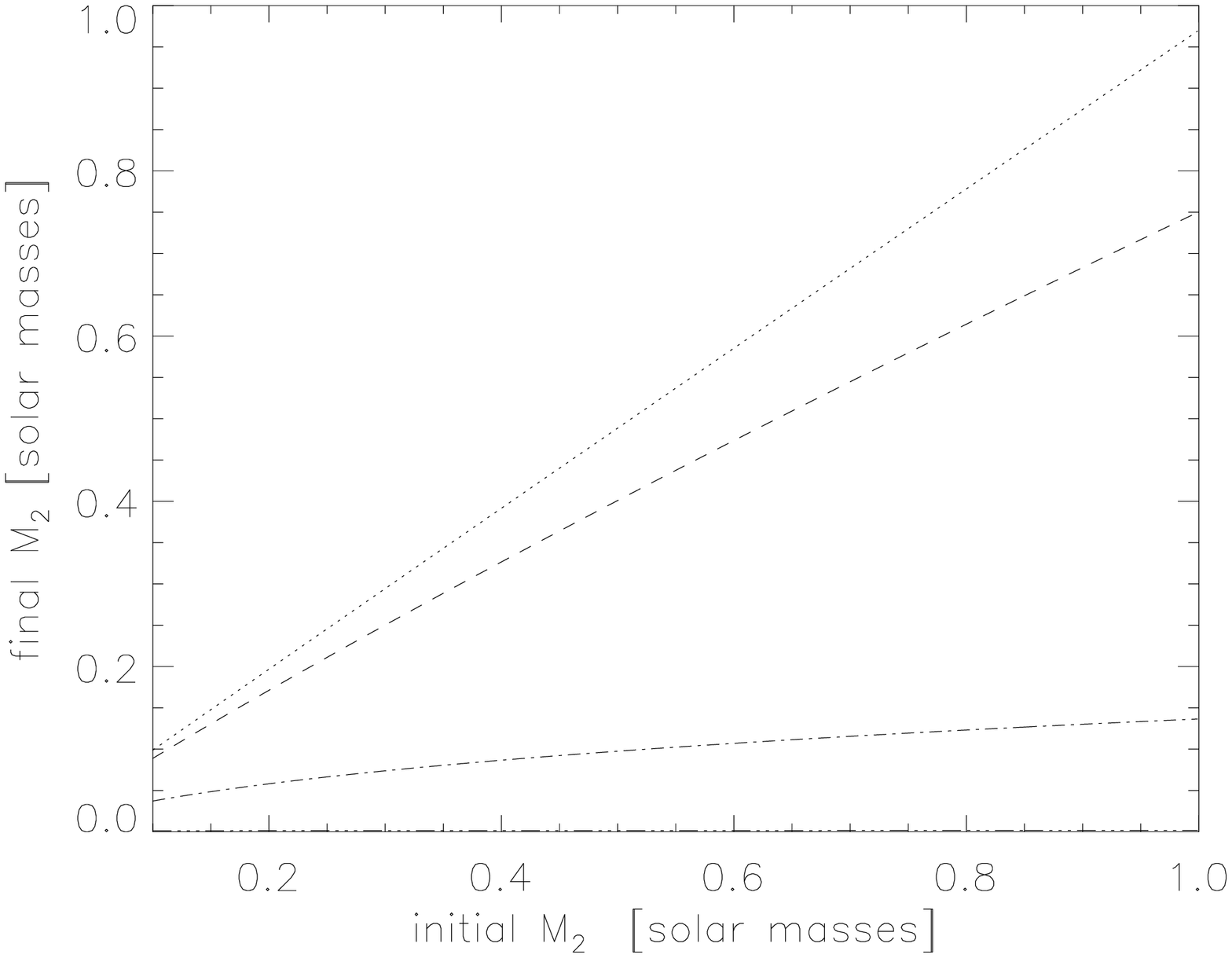} 
\vskip 10mm
\includegraphics[width = 90mm, angle = 90]{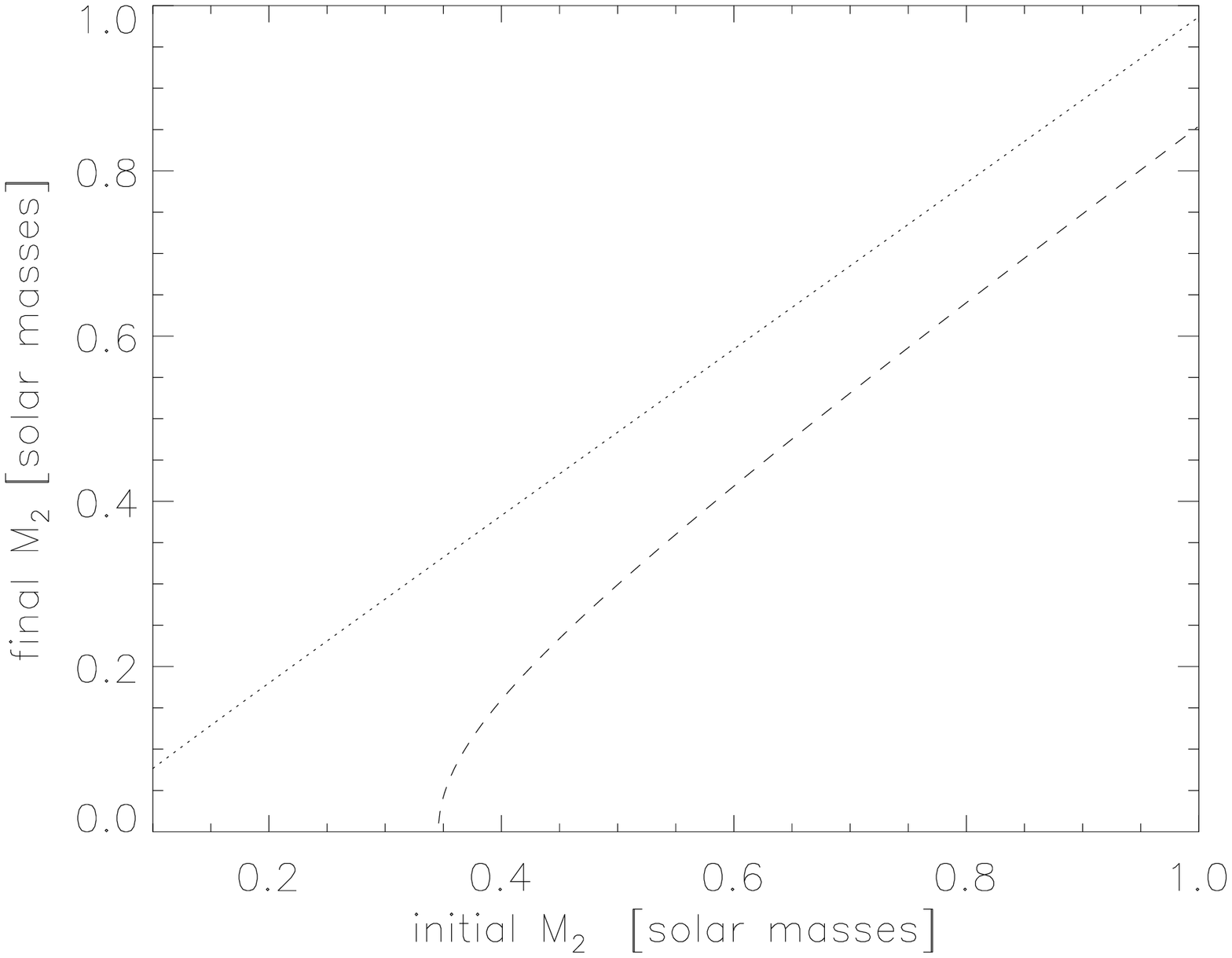} 
\end{center}
\caption{The final mass of the RD after the evaporation process 
from the mageto-dipole radiation from the NS versus the 
initial mass for different values of the efficiency $\eta$.
Upper panel: the radius of the secondary is always taken as the ZAMS
radius for a lower main sequence star. Lower panel: the radius of
the secondary is taken as the Roche radius at the evaporation onset.
Dotted line: $\eta=10^{-3}$; dashed line: $\eta=10^{-2}$, dot-dashed line:
$\eta=10^{-1}$.
In both cases, the orbital separation is fixed to~$1.5\times 10^{11}\cm$,
correspondent to an orbital period of $6.67\hr$ for a binary system with
$M_1=1.4\msun$ and $M_2=0.4\msun$.  
}
\label{f-evap}
\end{figure}

Figure~\ref{f-evap} shows the final mass of the companion star
versus its initial mass in these two limiting cases for several
values of the parameter $\eta$. The fixed parameters are the NS's initial
period ($3~\rm ms$) and the  orbital separation ($a=1.5\times 10^{11}\cm$),
although an increase of $a$ is expected on 
account of the mass lost by  the system. 

In both cases, the secondary's final mass is $M_2\lesssim 0.4\msun$
provided that $\eta < 10^{-1}$, for an initial 
mass~$M_2\lesssim 0.6\msun$.
For efficiencies $\eta\gtrsim 10^{-1}$ and initial masses
$M_2\lesssim 0.6\msun$ the secondary is completely
evaporated in the Roche lobe filling case. If the 
$M_2-R_2$~relation~(\ref{e-lms})
 is followed, $M_2$ drops below $0.1\msun$ if $\eta\gtrsim0.1$.
We may conclude that (in either scenario) the secondary is left with a mass
$M_2\lesssim 0.4\msun$ if its initial mass is $M_2\simeq 0.6\msun$
and the efficiency is $\eta\simeq 10^{-2}$; this value is 
consistent with the estimates $\eta\simeq 10^{-2}-10^{-1}$ 
cited in the literature \citep[e.g.][]{Tav92}.
The mass lost during the evaporation 
widened  the orbit by about~$\simeq 10\%$. (e.g. \citealp{Pos06}).

The evaporation scenario we have explored is perhaps the most unfavourable
for the survival of the companion, as some assumptions may be a little bit
unnecessarily  pessimistic. 
First of all, we have assumed that the relativistic wind is isotropic. This may
not be the case: in some models for the birth of magnetars (in connection
with the problem of long Gamma-ray Bursts)  most of the rotation energy is
removed by highly collimated jets \citep{Buc07c}.
If the collimated beam keeps away from the RD, the problem of evaporation
disappears.
Second, we assumed a fast initial NS spin period ($P_0=3\ms$), as 
required to build up the magnetar's field 
by a  $\alpha-\Omega$ dynamo \citep{Dun92,Tho93}. There are, however, 
alternative models which explain the formation of strong magnetic fields
as due to the amplification of high fossil fields due to flux conservation during the core collapse \citep{Fer05,Fer06}; in this case
there is no need of a very NS fast initial rotation period.
On account of the strong dependency of $L_{\rm SD} \propto P_0^{-2}$
(Equation~(\ref{e-sdl})), also the wind hitting the RD is less
energetic. 
To summarise the discussion relative to point~(1), there are several
possible scenarios in which the evaporation of the companion is not an issue.
Even in the worst case we have analysed in some detail, for reasonable
combinations of the parameters the RD may survive  evaporation.

Although the early relativistic wind is unable to evaporate the whole
star, it is not without consequences. Indeed, the
secondary star quickly recovers its hydrostatic equilibrium, but takes
a longer time  (see Equation~(\ref{e-kh})) to restore its thermal equilibrium. 
It is quite difficult to address the structure of the companion after it has 
been shaken  by the   magnetar wind. The companion is likely to be larger than
a ZAMS star with the same mass, 
so it may fill its Roche lobe 
and transfer mass to the NS,  powering (part of) its X-ray luminosity.

\bigskip

As far as query (2) is concerned, 
we expect that when the RD returns to thermal equilibrium,
it shrinks, fitting inside its own Roche lobe. Mass transfer ceases
and only the residual magnetar like emission remains.
This likely occurs on the Kelvin-Helmholtz time~(\ref{e-kh}).
Later on, we expect that also the magnetar field will decay.
Current theoretical models predict that the present field of~$10^{15}\G$
will decay to~$10^{12}\G$ (typical of the magnetic field of NSs in
ordinary LMXBs) in~$\lesssim 10^8\yr$  \citep{Hey98b,Col00, Gep02}
on account of non-linear field effects like ambipolar diffusion and Hall drift.
Later  on, accretion from the companion sets in that can induce further 
field decay, driving it  below $\simeq 10^{10}\G$ \citep{Bha91, Gep94}.
As the companion star will exhaust its nuclear
fuel in few~$10^9\yr$, it will swell up as a giant, filling its Roche lobe
and transferring mass to the NS. Since the NS's field is weak by now,
the accretion flow may settle in a disc. The ensuing spin up torque 
may accelerate the NS up to few milliseconds, and at the end of this
stage the system  would look like a {\em bona fide} LMXB.

A speculative deduction of our model is that at least some LMXBs 
may have systems  like  1E as progenitors.
The magnetic field of a NS born as a magnetar 
in a low mass binary might keep stronger than the field of a NS in a LMXB 
or of  a binary millisecond pulsar. 
This kind of higher field is observed in the seven 
Soft X-Ray Transients LMXBs where an X-ray pulsation is detected
during type I burst. This will be explored in a forthcoming paper.

\section{Summary}
\label{s-summary}

In this paper we have put forward a model for the unique X-ray source
1E161348-5055 (1E) an the centre of the young supernova remnant RCW~103. 
In our model 1E is a binary system made by a NS and a lower main
sequence star.  We interpret the signature of~$6.67\hr$ as the NS's spin
period. This requires that the NS is endowed with a strong
magnetic field of~$\sim 10^{15}\G$. 

The spin evolution of the NS depends on the relative importance of
the material and magnetic torques. 
If the magnetic torque is dominant
(which is the case if most of the X-ray luminosity is powered by magnetar
emission), the NS's rotation is almost synchronous with the orbital period,
similarly to what happens in Polar Cataclysmic Variables.
On the other hand, if the material torques are more relevant, then the 
spin period is about~$50-70\%$ of the orbital period. In this case, the 
system would be similar to an  Intermediate Polar
CV, with a NS instead of a white dwarf.
The magnetar-powered emission and the full orbital synchronisation 
scenario is most likely to occur if the secondary star is small
(i.e. significantly below the current upper 
limits derived by \citet{DeL06}).

Alternatively, an early relativistic
wind from the NS may have stirred an evaporation  wind from the companion.
On account of the large amount of energy injected into the RD's envelope,
the RD may be out of thermal equilibrium.
It may be larger than a main sequence star with the same mass, possibly  
powering  by accretion  the 
NS' X-ray luminosity. This phase is expected to last $10^4\yr$, the 
Kelvin-Helmholtz time scale of the RD  layers involved in the energy 
deposition from the early relativistic wind.  As soon as the star
recovers its thermal equilibrium, the star fits again within its own Roche 
lobe, and mass transfer will end. 

If the model we have suggested is correct, then 1E is the first case (to our knowledge) of
a magnetar hosted in a low mass binary system. This may spur  some interesting
developments in the studies of the evolution of narrow binary systems
harbouring  a NS.

\acknowledgements
It is a pleasure to thank Sergey Popov for useful discussion and Ed van den Heuvel for his insightful help.
This research has been supported by the Italian Ministry of University and
Research undre the grant PRIN 2005024090\_002.
We also thank an anonymous referee whose comments helped us to improve the
paper.

\clearpage
\nocite{}

\bibliographystyle{astron}
\bibliography{refs}


\clearpage
\appendix

\section{Equilibrium Solutions}
\label{s-equilibrium}

In this Appendix  we study in more detail the properties of the equilibrium
solutions found numerically in Section~\ref{s-polar}.
To this end it is convenient to cast Equation~(\ref{e-spin}) in a 
dimensionless form.
We measure the time in units of $\Omega_{\rm orb}^{-1}$, i.e. we set
$\tilde{t}=t\;\Omega_{\rm orb}$; we then introduce the synodic phases
\be
\phi_1' = \phi_1- \tilde{t},
\qquad
\phi_2' = \phi_2 - \tilde{t}.
\ee
If the secondary is synchronous, then $\phi_2'$ is independent of time.
After some algebra, Equation~(\ref{e-spin}) reads
\be
\label{e-denergy}
\frac{d \mathcal E}{d\,\tilde{t}} = \xi \; W(\xi)
\ee
where
\begin{eqnarray}
\label{e-energy1}
&\xi = \dot{\phi_1'} 
\\
\label{e-energy2}
&\mathcal E = \frac{1}{2}\;\xi^2 + n_{\rm dd} \;
( \sin\phi_2'\:\sin\phi_1' - 2\,\cos\phi_2'\:\cos\phi_1')
\\
\label{e-work}
&W(\xi) = -n_{\rm md} \; (1+\xi)^3 - n_{\rm diss}\; \xi
+ n_{\rm ap} \; \left[\frac{1}{1 + \lambda \: (1+\xi)^2}  - 
\frac{\lambda}{2}  \:   (1+\xi)^2\right],
\end{eqnarray}
and
\begin{eqnarray}
&n_{\rm md} = \twothirds \frac{\mu_1^2\;\Omega_{\rm orb}}{I_1\;c^3}
\\
&n_{\rm diss} = 1/(\Omega_{\rm orb}\;\tau_{\rm syn})
\\
&n_{\rm ap} = \frac{|\dot{M}_2| \: (G\: M\; b_1)^{1/2}}{I_1\;\Omega^2_{\rm orb}}
\\
&n_{\rm dd} = \frac{\mu_1\;\mu_2}{G\: M \:I_1}
\end{eqnarray}
are the dimensionless torques on the NS.
Equation~(\ref{e-denergy}) has the form of an energy equation in a 
non-conservative system, on account of the forcing term  
$\xi \; W(\xi)$ on its right-hand side.
The dipole-dipole torque does not appear in the braking function $W$,
showing that this torque does not help to attain synchronism, 
since it does not alter the ``energy'' $\mathcal E$.

If there is no mass transfer, then $n_{\rm ap}=0$, and since
$n_{\rm md}\ll n_{\rm diss}$ Equation~(\ref{e-denergy}) reads approximately
\be
\frac{d \mathcal E}{d\,\tilde{t}} \simeq - n_{\rm diss}\; \xi^2 
\ee
Starting from a non synchronous configuration ($\xi\neq 0$), the 
right-hand side of this equation is negative, so  
$d\mathcal E/d\tau<0$. The system then evolves towards the minimum of 
$\mathcal E$, given by  $\xi\equiv\dot{\phi}_1'=0$ and
\be
\label{e-angle}
\cos\phi_1' =  \frac{2\: \cos \phi_2'}{\sqrt{1+3\cos^2\phi_2'}}
\qquad
\sin\phi_1' = -\frac{\sin \phi_2'}{\sqrt{1+3\cos^2\phi_2'}}.
\ee
This is the synchronous solution, where the synodic phase $\phi_1'$ is locked
to the angle defined by the last couple of equations.
It is also interesting  to note  that the "potential"
\be
V(\phi_1') = 
n_{\rm dd} \; (\sin\phi_2'\; \sin\phi_1' - 2\,\cos\phi_2'\;\cos\phi_1')
\ee
in Equation~(\ref{e-energy2}) is proportional
to the the magnetic interaction potential energy between  the stellar
 magnetic dipoles $U=-\muns\cdot\bb_2(1)$. The equilibrium synodic 
position angle of the synchronous solution 
therefore  minimises  the interaction
magnetic potential energy between the two stars. 

The qualitative pattern of the evolution towards synchronism does not
change if $W$  remains negative. The only positive term
in $W$ which may change the sign of $W$ and alter this conclusion is the
first term in square brackets proportional to $n_{\rm ap}$. This is
the accretion term, which tends to spin up the NS. We now show that
for a large accretion rate, the NS tends to a stable asynchronous
uniform rotation.  If $n_{\rm ap}$ exceeds a threshold value, 
the equation  $W(\xi)=0$ 
has a root $\xi_0$. Since $W'(\xi)<0$ for any $\xi\geq0$, 
this root is also unique.  In addition,
$W(\xi) > 0$  if $\xi < \xi_0$ and $W(\xi) < 0$ if $\xi > \xi_0$.
These results show that $d\mathcal E/d{\tilde t} <0$ for $\xi > \xi_0$
and  $d\mathcal E/d{\tilde t} >0$ for $\xi < \xi_0$: 
the system evolves towards the value of $\mathcal E$ characterised by
$\xi = \xi_0$ and  $\chi = \xi_0 \; \tilde{t} + \phi_0$.
Since $\xi_0\neq 0$, this is an asynchronous solution:
the system cannot attain synchronism when the 
accretion spin up torque exceeds a critical value, in agreement with 
what is well known in the literature for Polar
systems (see e.g. \citealp{Cam86a}).

\begin{figure}
\begin{center}
\includegraphics[width =90mm, angle = 90]{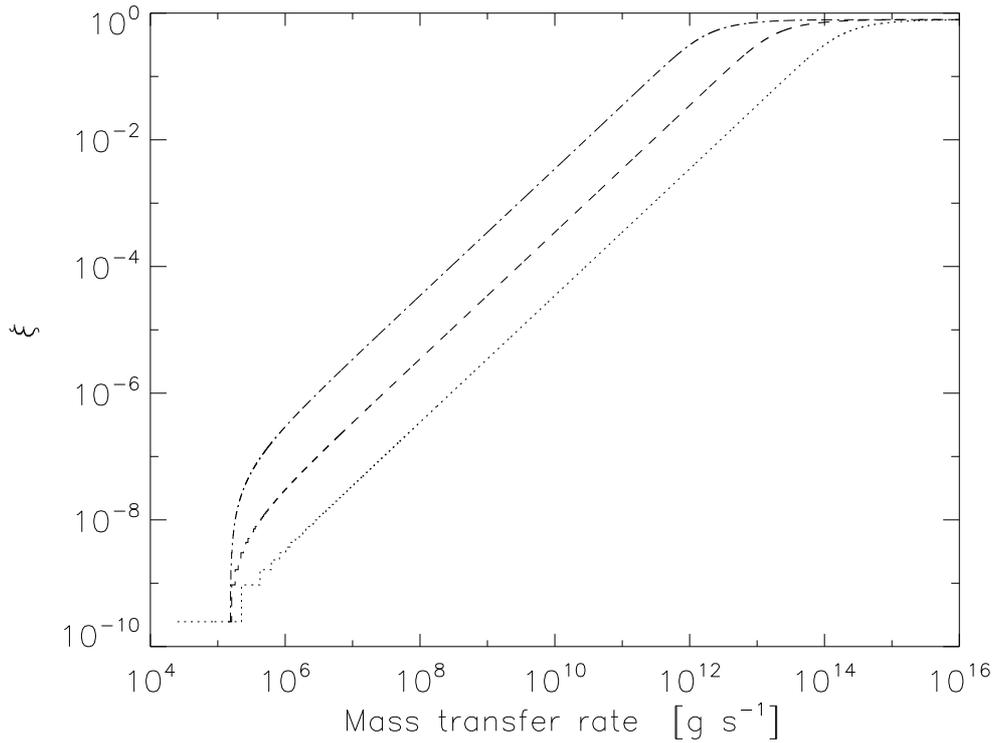}
\end{center}
\caption{The normalised synodic equilibrium  frequency 
$\xi=(\omega_1-\Omega_{\rm orb})/\Omega_{\rm orb}$ plotted against the mass 
transfer rate  $|\dot{M}_2|$ from the secondary star, for different values
of the synchronisation time $\tau_{\rm syn}$.
Dotted line, $\tau_{\rm syn}=10\yr$;
dashed line  $\tau_{\rm syn}=100\yr$
and dot-dashed line $\tau_{\rm syn}=1000\yr$.
As $|\dot{M_2}|$ increases, the value of $\xi$ tends asymptotically to the
value $\xi\simeq 0.7$.}
\label{f-async}
\end{figure}

The result is shown in  Figure~\ref{f-async}, 
which  plots the equilibrium synodic frequency $\xi$
against the mass transfer rate $|\dot{M}_2|$, for different values of the
synchronisation time $\tau_{\rm syn}$. 
As expected, $\xi$ grows with $|\dot{M}_2|$. Below the  critical
threshold $|\dot{M}_2|\sim 10^5\g\s^{-1}$ 
the equilibrium solution  is always synchronous. As the dissipative
torque decreases (i.e. for long  $\tau_{\rm syn}$), synchronism is more
difficult to achieve.  For $|\dot{M}_2|\simeq 10^{13}\g\s^{-1}$
(the mass accretion rate inferred from the source's X-ray luminosity)
and the most favourable case  $\tau_{\rm syn}=10\yr$ we find that
the NS has a degree of asynchronism of a few percent.


\clearpage

\end{document}